\documentclass[12pt]{article}
\usepackage[cp1251]{inputenc}
\usepackage{amsmath,amssymb}

\usepackage[dvips]{epsfig}
\usepackage[dvips]{graphics}
\begin{document}
%\twocolumn[ \Arthead{6}{2005}{4 (24)}{1}{10}

\begin{center}
\Large\textbf{Torsion and gravitational interaction in Riemann-Cartan space-time}\\
\bigskip
\normalsize A.V. Minkevich\\
\medskip
\textit{Department of Theoretical Physics and Astrophysics, Belarusian State University,
Minsk, Belarus\\} \textit{Department of Physics and Computer Methods, Warmia and
Mazury University in Olsztyn, Olsztyn, Poland\\}
 %E-mail:MinkAV@bsu.by; awm@matman.uwm.edu.pl
\end{center}

\begin{center}
\begin{minipage}{0.8\textwidth}
\textbf {Abstract.} The influence of space-time torsion on gravitational interaction at
cosmological and astrophysical scales is discussed within the framework of gauge gravitation
theory in Riemann-Cartan space-time. It is shown that the interaction of the vacuum torsion
with proper angular momentums of gravitating objects leads to appearance of additional
gravitational force which can be manifested at astrophysical scale.
\end{minipage}
\end{center}

\section{Introduction}

The gauge gravitation theory in 4-dimensional Riemann-Cartan spacetime $U_4$ (GTRC) is
a necessary generalization of metric gravitation theory in the framework of gauge approach
by including the Lorentz group into the gauge group corresponding to gravitational
interaction. Gravitational equations of GTRC and their physical consequences depend on the
choice of gravitational Lagrangian $\mathcal{L}_{\rm g}$ as function of gravitational field
strengths - the curvature $F^{ik}{}_{\mu\nu}$ and torsion $S^i{}_{\mu\nu}$ tensors, and
also on the coupling of matter with gravitational field. By using minimal coupling the
energy-momentum and spin momentum tensors of gravitating matter play the role of sources
of gravitational field. Pioneer works dedicated to GTRC were connected with investigation of
Einstein-Cartan theory, gravitational Lagrangian of which is given in the form of scalar
curvature of $U_4$ \cite{k1,k2,k3} (see also \cite{k4,k5}) \footnote{An important
contribution to research of Einstein-Cartan theory was made by polish physicists (A.
Trautman, W. Kopczynski, B. Kuchowicz, J. Tafel) in connection with investigation of the
problem of cosmological singularity.}. In the frame of Einstein-Cartan theory the torsion
tensor is linear algebraic function of spin momentum of gravitating matter and in the case of
spinless matter the influence of torsion on dynamics of gravitating system vanishes. In
connection with this the opinion that the torsion is created by spin momentum of matter and
in the case of spinless matter has to vanish is widely extended in literature. However, such
situation indicates an exceptional position of Einstein-Cartan theory. Really by taking into
account that the torsion tensor is gravitational field strength corresponding to 4-translations
subgroup of gauge group, which according to Noether theorem is connected with
energy-momentum tensor, we have to conclude that generally torsion in the frame of GTRC
can be created by energy-momentum tensor. Such situation takes place in the frame of
GTRC based on gravitational Lagrangians including quadratic in the curvature and torsion
terms \footnote{First this was pointed in the case of homogeneous isotropic models (HIM)
built in \cite{m1} and in series of papers dedicated to GTRC \cite{h}.}. In the frame of
gauge approach the Lagrangian of gauge field usually is given as function quadratic in the
gauge field strength; existence of many invariants quadratic in the curvature and torsion
tensors is notable feature of GTRC, moreover there is linear in the curvature invariant - scalar
curvature. Because the detailed form of gravitational Lagrangian is unknown, we will consider
GTRC based on the following sufficiently general expression of $\mathcal{L}_{\rm g}$
used in a number of papers\footnote{The definitions and notations of our previous papers
(see e.g. \cite{m2}) are used below. With the purpose to make quantitative estimations the
light velocity $c$ is conserved in formulas.}:
\begin{eqnarray}\label{1}%\fl
\mathcal{L}_{\rm g}=  f_0\,
F+F^{\alpha\beta\mu\nu}\left(f_1\:F_{\alpha\beta\mu\nu}+f_2\:
F_{\alpha\mu\beta\nu}+f_3\:F_{\mu\nu\alpha\beta}\right)  %\nonumber \\
+ F^{\mu\nu}\left(f_4\:F_{\mu\nu}  +f_5\: F_{\nu\mu}\right)
    \nonumber \\
+ f_6\:F^2 %\nonumber \\
+S^{\alpha\mu\nu}\left(a_1\:S_{\alpha\mu\nu}+a_2\: S_{\nu\mu\alpha}\right)
+a_3\:S^\alpha{}_{\mu\alpha}S_\beta{}^{\mu\beta},%\nonumber
\end{eqnarray}
where $f_i$ ($i=1,2,\ldots,6$), $a_k$ ($k=1,2,3$) are indefinite parameters,
$f_0=\frac{c^{4}}{16\pi G}$ ($G$ is Newton's gravitational constant)\footnote{In the case
of GTRC, which is not invariant with respect to transformations of spatial inversions, a
quantity of additional invariants can be built by using Levi-Civita discriminant tensor and
added with indefinite parameters to $\mathcal{L}_{\rm g}$ (1). Such theories were studied
in a number of papers (see e.g. \cite{m3,m4,hehl}).}. If one supposes that GTRC
corresponds to real Universe, we have to determine values of parameters $f_i$ and $a_k$ in
expression (1). Restrictions on indefinite parameters of $\mathcal{L}_{\rm g}$ can be found
on request that GTRC allows to solve some principal problems of general relativity theory
(GR) and physical consequences of this theory are the most satisfactory. Some such
restrictions were found from analysis of isotropic cosmology built in the frame of GTRC
based on $\mathcal{L}_{\rm g}$ (1). It should be noted that in the case of spatially
homogeneous isotropic matter contribution assuming in the frame of isotropic cosmology the
average of spin momentum is equal to zero. The cosmological equations generalizing
Friedmann cosmological equations of GR and equations for torsion functions were given in
\cite{m5} (see also \cite{m2}) in general form without using any restrictions on parameters
$f_i$ and $a_k$. The investigation of these equations leads to the following restrictions:
\begin{eqnarray}\label{1.2}
2a_1 + a_2  +3a_3=0, \qquad   2f_1 - f_2=0,
\end{eqnarray}
by which the solution of the problem of cosmological singularity and the dark energy
problem was obtained \cite{m5,m6}. Then equations of isotropic cosmology include three
indefinite parameters: parameter $\alpha=\frac{f}{3f_0^2}$  ($f = f_1  + \frac{{f_2 }} {2} +
f_3 + f_4 + f_5 + 3f_{6}>0$) with inverse dimension of energy density, parameter $b=a_2 -
a_1$ with the same dimension as $f_0$ and dimensionless parameter $\omega= \frac {f_2 +
4f_3 + f_4 + f_5} {f}$. The solution of indicated cosmological problems together with
fulfillment of the correspondence principle with GR leads to the following restrictions, which
were defined more exactly by analysis of gravitational equations of GTRC \cite{m8,m9}:
\begin{eqnarray}\label{1.3}
 0<x=1-\frac{b}{f_0}\ll 1, \qquad  0 < \omega\ll 1
\end{eqnarray}
and the value of parameter $\alpha^{-1}$ corresponds to some high energy density. There is
a number of papers dedicated to isotropic cosmology with other restrictions on indefinite
parameters (see e.g. \cite{n1,n2,xi}) given in accordance with analysis of particle content of
linearized GTRC fulfilled by supposition that physical space-time in the vacuum is
Minkowski space-time \cite{h}. However, as it was shown in \cite{m5} on the base of
analysis of equations of isotropic cosmology the physical space-time in the vacuum in the
frame of GTRC has the structure of Riemann-Cartan continuum with de Sitter metric and the
strict analysis of particle content has to be connected with consideration of gravitational
perturbations above the vacuum space-time of GTRC. It should be noted that the deviation
of the structure of the vacuum space-time in the frame of GTRC from Minkowski
space-time, which is essential at cosmological scale, can be unimportant by local analysis
given in \cite{h} because of smallness of corresponding characteristics of metric and torsion
for the vacuum (see below). However, we have to consider corresponding results of \cite{h}
as approximative whose range of applicability is limited by weak fields.

The space-time torsion plays the principal role by the change of gravitational interaction by
certain conditions. Unlike isotropic cosmology, where space-time torsion is created by
spinless matter and spin momentum of gravitating matter is not demonstrated, the interaction
of torsion with spinning matter can play principal role in astrophysics (galaxies, galactic
clusters).

Isotropic cosmology in the frame of GTRC based on gravitational Lagrangian (1) was
investigated in a number of our papers (see e.g. \cite{m2,m5,m6,m7,m8,m9,m10} and Refs.
herein). In Section 2 some relations of isotropic cosmology are given in connection with
consideration of the role of space-time torsion that is used by discussion of the influence of
torsion on gravitational interaction at astrophysical scale in Section 3.

\section{Gravitational interaction at cosmological scale and vacuum torsion}

Any HIM in Riemann-Cartan space-time is described by three functions of time: the scale
factor of Robertson-Walker metric $R(t)$ and two torsion functions - scalar function
$S_{1}(t)$ and pseudoscalar function $S_{2}(t)$. Cosmological equations generalizing
Friedmann cosmological equations of GR by using restrictions (2) take the form \cite{m6}
\begin{eqnarray}\label{2.1}%\fl
    \frac{k}{R^2} + (H-2S_1)^2 -S_2^2= \nonumber\\
    \frac{1}{{6f_0 Z}}
        \left[
            {\rho c^2  -6 b S_2^2
            + \frac{\alpha }{4} \left( {\rho c^2  - 3p - 12bS_2^2 } \right)^2 }
        \right],
\end{eqnarray}
\begin{eqnarray}\label{2.2}%\fl
    \dot{H}-2\dot{S}_1 +H (H-2S_1)= \nonumber\\
    -\frac{1} {{12f_0 Z}}
        \left[
            \rho c^2  + 3p - \frac{\alpha } {2} \left( {\rho c^2 - 3p - 12bS_2^2 } \right)^2
        \right],
\end{eqnarray}
where $H=\dot{R}/R $ is the Hubble parameter (a dot denotes the differentiation with
respect to $x^0= c t$), $k=+1,0,-1$ for closed, flat and open models respectively, $\rho$ is
mass density, $p$ is pressure and $Z=1+\alpha\left( \rho c^2 - 3p - 12b S_2^2\right)$. The
torsion functions $S_1$ and $S_2$ are:
\begin{eqnarray}\label{2.3}%\fl
    S_1  = -\frac{\alpha }{4Z} [\dot \rho c^2
    - 3 \dot p + 12f_0 \omega H S_2^2
    -12( {2b - \omega f_0 } ) S_2 \dot S_2],
\end{eqnarray}
\begin{eqnarray}\label{2.4}
 S_{2}^{2}  = \frac{\rho c^2 - 3p}{12b} + \frac
{1-(b/2f_0) (1 +  \sqrt{X})} {12b \alpha (1- \omega/4)},
\end{eqnarray}
where
\begin{equation}\label{2.5}
X=1+ \omega (f_0^2/b^2) [1- (b/f_0) - 2(1- \omega /4) \alpha ( \rho c^2+ 3p)]\ge
0.
\end{equation}
In accordance with gravitational equations of GTRC functions $\rho$ and $p$ satisfy the
equation as in GR:
\begin{equation}\label{2.6}
\dot \rho + 3H(\rho + p/c^2)=0.
\end{equation}
By given equation of state of gravitating matter eqs. (4)-(7) allow to find cosmological
solutions of GTRC. Behaviour of cosmological solutions in GTRC differs essentially from
that of GR at the beginning of cosmological expansion and at asymptotics, where the torsion
plays the important role. This follows directly from expression (7) for torsion function
$S_{2}^{2}$.  The presence of $\sqrt{X}$ in formula (7) leads to appearance of limiting
(i.e. maximum allowable) energy density. Gravitational interaction near limiting energy density
has repulsive character and corresponding cosmological solutions describe regular transition
from compression to expansion ("Big Bounce"). In the case of HIM with restrictions (3)
filled with gravitating matter with equation of state $p=p(\rho)$ the Hubble parameter with its
time derivative near a bounce in the first approximation with respect to $\sqrt{X}$ are:
\begin{eqnarray}\label{2.7}
& & H_{\pm}=\pm \frac{2b^2}{3f_0^2 \omega \alpha} \frac{\sqrt{X} [(1/4b)(\rho_m c^2
+p_m) - (k/R^2) - \frac{1-b/(2f_0)}{24f_0 \alpha}]^{1/2}}{(3\frac{1}{c^2}\frac{d p_m}{d\rho_m}+1 ) (\rho_m c^2+p_m)},
\nonumber\\
& & \dot{H}=\frac{4b^2}{3f_0^2\omega \alpha } \frac{(1/4b)(\rho_m c^2 +p_m) -
(k/R^2) - \frac{1-b/(2f_0)}{24f_0 \alpha}}{(3\frac{1}{c^2}\frac{d p_m}{d\rho_m}+1 ) (\rho_m c^2+p_m)}.
\end{eqnarray}
$H_{-}$- and $H_{+}$-solutions describe the stages of compression and expansion
correspondingly, and the transition from compression to expansion takes place by reaching
limiting energy density determined from equality $X=0$: $(\rho_{max}c^2) \sim
(\omega\alpha)^{-1}$, which in the frame of our classical theory has to be less than the
Planckian one. Because energy density near a bounce is close to $(\rho_{max}c^2)$ , the
constant term $\frac{1-b/(2f_0)}{24f_0 \alpha}$ in (10) is not essential.

The principal influence of torsion on gravitational interaction becomes apparent also when
energy density is small and its influence on geometrical structure of space-time in the vacuum
is essential. Unlike GR (without cosmological constant) where space-time in the vacuum (in
the case of flat models with $k=0$) is Minkowski space-time,  in the frame of GTRC
space-time in the vacuum has the structure of Riemann-Cartan continuum with de Sitter
metric \cite{m5}. It is connected with the presence of constant term - vacuum torsion - in
expression (7) of $S_{2}^{2}$:
\begin{equation}\label{2.8}
S_{2}^{2(vac)}= \left [1 - \frac{b}{2f_0}[1 + \left(1- \omega (1-b/f_0) \frac{f_0^2}{b^2}
\right)^{1/2}]\right ][12\alpha b(1 - \omega/4)]^{-1}.
\end{equation}
Then in accordance with eqs. (4)-(7) the vacuum value of $H^2$ (in the case $k=0$) is:
\begin{equation}\label{2.9}
H^{2(vac)}= \frac{6b^2} {f_0} \alpha S_2^{4(vac)} [1 - 6 \alpha (2b +\omega f_0) S_2^{4(vac)}]^{-1}.
\end{equation}
At asymptotics when energy density is small ($\alpha \rho c^2 \ll 1$), by using the restriction
$ 0 < x=1-\frac{b}{f_0} \ll 1 $  the expression (7) for $S_{2}^{2}$ in the lowest
approximation with respect to $x$ takes the form:
\begin{equation}\label{2.10}
S_2^2  = \frac {1} {12b} \left[\rho c^2  - 3p + \frac {1  - b/f_0} {\alpha}\right],
\end{equation}
and as a result cosmological equations (4)-(5) at asymptotics are:
\begin{equation}\label{2.11}
    \frac{k}{R^2 } + H^2  = \frac{1}{6b }\left[\rho c^2 + \frac{1}{4\alpha} \left(1 - \frac{b}{f_0}\right)^2
     \right],
\end{equation}
\begin{equation}\label{2.12}
    \dot H + H^2  =  - \frac{1} {{12b }}\left[ (\rho c^2 + 3p) - \frac{1}{2\alpha}
    \left(1 - \frac{b}{f_0}\right)^2 \right].
\end{equation}
According to (13)-(14) and in compliance with (11)-(12) we have:
\begin{equation}\label{2.13}
S_{2}^{2(vac)}  =  \frac {1  - b/f_0} {12b \alpha}, \qquad  H^{2(vac)}=\frac{\left(1 - \frac{b}{f_0}\right)^2}{24b\alpha }.
\end{equation}
The effective cosmological constant in (14)-(15) is induced by the vacuum torsion
$S_{2}^{2(vac)}$. Unlike standard $\Lambda CDM$-model effective cosmological constant
appears in (14)-(15) as a result of solution of gravitational equations for HIM that leads to the
change of gravitational interaction when energy density is small and comparable with
cosmological constant - the vacuum gravitational repulsion effect leading to accelerating
cosmological expansion at present epoch.

Because of restriction $ 0<x=1-\frac{b}{f_0}\ll 1$  the vacuum value of $H$ and the vacuum
torsion function $|S_1|$ are negligibly small in comparison with $|S_{2}^{(vac)}|$. Owing to
this the curvature tensor (see \cite{m7,m2}) has the following vacuum components:
\begin{equation}
F^{12}{}_{12} = F^{13}{}_{13} = F^{23}{}_{23} = - S_{2}^{2(vac)}.
\end{equation}
Unlike the torsion function $S_1$, the influence of which on gravitational interaction is
essential only at extreme conditions near a bounce, the torsion function $S_{2}^{2}$ plays
important role also at asymptotics. Connection between parameters $b$ and $\alpha$ can be
found by supposition that the value of effective cosmological constant in eqs. (14)-(15)
corresponds to observable accelerating cosmological expansion. Parameter $\omega$
together with $\alpha$ play important role at extreme conditions near a bounce. By energy
densities, which are much smaller than limiting energy density and greater than constant term
in (14) $ \frac{1}{4\alpha} \left(1 - \frac{b}{f_0}\right)^2$ the behaviour of cosmological
solutions of eqs. (14)-(15) practically coincides with that of Friedmann cosmological
equations of GR.We deal with such energy densities in astrophysics in the case of various
objects in galaxies and galactic clusters. In general case the description of such systems in
the frame of GTRC is difficult problem because of complexity of gravitational equations.
The situation is simplifying, when minimum GTRC was determined \cite{m9}. This theory
includes three indefinite parameters, which can be expressed through parameters of isotropic
cosmology \footnote{The gravitational Lagrangian of minimum GTRC is given by (1), if we
assume that $f_1=f_2=f_3=f_4=0$, $a_1=b, a_2=2b, a_3=-\frac{4}{3}b, f_5=3f_0^2
\alpha \omega, f_6=f_0^2 \alpha (1-\omega)$ and $0 < \omega\ll 1$.}. By neglecting terms
with small parameter $\omega$ equations of minimum GTRC lead to gravitation equations
for metric in the form of Einstein gravitation equations with cosmological constant, which are
valid for spinless gravitating systems at wide range of energy density - beginning with
extremely high energy densities defined by $\alpha^{-1}$:
\begin{equation}\label{2.15}
G^{\mu}{}_{\lambda}=-\frac{1}{2b}
 \left[T_{\lambda}{}^{\mu} + \delta^{\mu}_{\lambda}
\frac{(1-\frac{b}{f_0})^2}{12\alpha} \right],
\end{equation}
where $G^{\mu}{}_{\lambda}$ is Einstein tensor. The influence of torsion appears in eq.
(18) via formation of effective cosmological constant and the change of gravitational
constant. We see that the correspondence principle with GR will be fulfilled if parameter $b$
satisfies the condition $0<1- \frac{b} {f_0}\ll 1$. Nonvanishing components of torsion
tensor $S_{\alpha\mu\nu}$ satisfy the relation \cite{m9}:
\begin{equation}\label{2.16}
S_{\lambda\mu\nu }(S^{\lambda\mu\nu }- 2
 S^{\mu\nu\lambda})= \frac {1} {2b} \left[T + \frac {1  - b/f_0} {\alpha}\right],
\end{equation}
where $T=T_{\mu}{}^{\mu}$. This relation corresponds to formula (13) of isotropic
cosmology, if we take into account that $S_2^2=-\frac{1}{6}
S_{\alpha\mu\nu}S^{\alpha\mu\nu}$ with $\alpha,\mu,\nu =1,2,3$ ( ${\alpha} \neq {\mu}$,
${\alpha} \neq {\nu}$).

\section{Vacuum torsion and gravitational interaction at astrophysical scale}

By neglecting spin effects the description of various astrophysical objects in the frame of
minimum GTRC practically coincides with that in GR \footnote{We don't discuss here
objects with extremely high energy densities, consideration of which is possible by taking
into account terms with parameter $\omega$ in gravitational equations.}, because the
influence of cosmological constant in (18) is negligibly small at astrophysical scale. In the
case of spinning matter corrections of the metric determined by eq. (18) are also sufficiently
small. However spin effects can be manifested  as a result of interaction of vacuum torsion
$S_{2}^{2(vac)}$ defined according to (16) with proper angular momentums of
astrophysical objects (stars in galaxies, galaxies in galactic clusters) that can have an influence
on their movement. Because the value of vacuum torsion is much greater than cosmological
constant (by virtue of restriction $ 0<x=1-\frac{b}{f_0}\ll 1$), its influence quantitatively can
become apparent at non-relativistic approximation.

With the purpose to study movement of an object with proper angular momentum in
gravitational field we will use equations of motion of particle with momentum in
Riemann-Cartan space-time \cite{ms} generalizing Papapetrou's equations for rotating
particle in GR \cite{p} \footnote {In \cite{ms} the curvature tensor was defined with
opposite sign and signature (+2) was used.}. In the case of rotating particle with angular
velocity tensor $\Omega_{ik}$ corresponding equations of motion by conserving terms,
which are essential at non-relativistic approximation, are:
\begin{equation}\label{3.1}
\frac {D P_i} {d \tau} = \frac{1}{2} I \Omega_{mn} F^{mn}{}_{il} v^{l} \qquad (i,l,m,n=1,2,3),
\end{equation}
where $\frac{D}{d \tau}$ denotes riemannian absolute derivative with respect to proper time
$\tau$, $P_i$ is generalized momentum, $I$ is inertia momentum  and $v^{l}$ is velocity of
particle. In non-relativistic approximation $P_i= m v_{i}$ ($m$ is particle mass),
$\Omega_{mn}=const$  and influence of curvature tensor in right side of (20) can become
apparent by means of vacuum curvature (17). The  right side of equation (20) determines
additional gravitational force connected with interaction of vacuum torsion with proper
angular momentum of particle.

As example we will consider the circular motion of rotating particle in spherically symmetric
gravitational field created by mass $M$ in non-relativistic approximation. By taking into
account that $g_{00}= 1+\frac{2\phi}{c^2}$ ($\phi$ is newtonian potential), components of
angular velocity $\Omega_{i}= \epsilon_{ikl} \Omega^{kl}$  and relation (17) we obtain in
the case of motion in plane $XOY$ (centrum of mass $M$ is in origin of coordinates, vector
of orbital angular momentum is directed along the axe $OZ$) equation of motion in usual
form $m \frac {d\mathbf{v}} {dt} = \mathbf{F}$ with the following expression of the force
vector:
\begin{equation}\label{3.2}
\mathbf{F}=-m\frac{d\phi}{d\mathbf{r}} + I \Omega_3 S_{2}^{2(vac)} v \frac{\mathbf{r}}{r}.
\end{equation}
The force (21)  includes besides Newtonian term additional force, direction of which
depends on relative orientation of proper and orbital angular momentums. We have the force
of attraction or repulsion depending on  $\Omega_3 < 0$ or $\Omega_3 > 0$, as result its
value is:
\begin{equation}\label{3.3}
F=G\frac{mM}{r^2} \pm I \Omega S_{2}^{2(vac)} v ,
\end{equation}
where $\Omega= |\Omega_3|$. By taking into account that the force (22) is centripetal force
we obtain the following dependence of velocity on distance from centrum and parameters of
particle and gravitational field:
\begin{equation}\label{3.4}
v=\pm \frac{I}{2m} \Omega S_{2}^{2(vac)} r + \left[(\frac{I}{2m} \Omega S_{2}^{2(vac)} r)^2 + \frac{GM}{r}\right]^{\frac{1}{2}}.
\end{equation}
By given parameters of particle and gravitational field values of the force (22) and velocity
(23) depend on parameter $x=1- \frac{b} {f_0}$. By taking into account that average mass
density in the Universe at present epoch $\rho_1 =\frac{x^2}{4c^2 \alpha}$ is of order
$10^{-26} \frac{\text{kg}}{\text{m}^3}$, we obtain that at the first approximation
\begin{equation}\label{3.5}
S_{2}^{2(vac)}=\frac{16 \pi G}{3c^2 x} \rho_1 \sim {\frac{10^{-52}}{x}} (\text{m}^{-2}).
\end{equation}
Now we will consider the application of (23) in the case of attraction force to star similar to
Solar ($I/m\sim10^{18}$ $\text{m}^2$, $\Omega \sim 0.5\cdot10^{-6}$ $\text{s}^{-1}$)
moving in galaxy similar to Andromeda ($M = 2\cdot10^{41}$ $\text{kg}$) by taking $x
=10^{-25}$ and consequently $S_{2}^{2(vac)}= 10^{-27} (\text{m}^{-2})$ that
corresponds to high energy density scale $\alpha^{-1} = 10^{7} \rho_{nucl}c^2$
($\rho_{nucl}$ is nuclear mass density). As numerical analysis shows at distances $r < 9$
kpc ($1 $\text{kpc}$=0,3086\cdot10^{20} \text{m}$) Newtonian term in (23) plays the
definitive role, by growth of $r$ from 9 kpc to 25 kpc the velocity $v$ according to
Newtonian law decreases from $219\cdot10^{3}$ km/s to $132\cdot10^{3}$ km/s, but
according to (3.4) the velocity $v$ changes only from $256\cdot10^{3}$ km/s to
$259\cdot10^{3}$ km/s. By further increase of $r$ essential growth of velocity $v$ takes
place according to (23); this effect can be essential in galactic clusters, where we deal with
vast space scale of order 10 $\text{Mpc}$ and more. We see that the force of interaction of
the vacuum torsion with proper angular momentums of stars can be essential by formation of
rotation curves in galaxies that can be interesting in connection with dark matter problem.
Effects discussed above at galactic scale (at $x \sim 10^{-25}$) are negligible when moving
the planets in the Solar system due to the smallness of the additional force in (22) in
comparison with the Newtonian force. For example, in the case of Earth the additional force
of attraction is only $10^{-12}$ part of the Newtonian force.

Effects connected with interaction of vacuum torsion  with proper angular momentums can
be important in astrophysics also in the case of systems of stars with high angular velocity of
proper rotation, for example systems of double pulsars.

Although given above consideration was realized in the frame of minimum GTRC, similar
effects take place in other GTRC because of existence of the vacuum torsion.

\section{Conclusion}

Research of gravitation theory in Riemann-Cartan space-time shows that satisfying the
correspondence principle with general relativity theory GTRC leads to certain principal
differences concerning gravitational interaction at cosmological and astrophysical scales.
Distinctions are connected with geometrical structure of physical space-time,  namely with
space-time torsion. The torsion created by spinless matter changes the character of
gravitational interaction at extreme conditions and leads to possible existence in the nature of
limiting energy density. Gravitational repulsion at extreme conditions ensures the regular
behaviour of all HIM, including inflationary cosmological models. The deviation of the
structure of physical space-time in the vacuum from that of Minkowski space-time leads also
to important physical consequences concerning gravitational interaction at cosmological and
astrophysical scales. The vacuum torsion generates effective cosmological constant by
changing gravitational interaction at cosmological scale when energy density is small that
allows to explain accelerating cosmological expansion at present epoch. The interaction of
the vacuum torsion with proper angular momentums of gravitating objects leads to
corrections of gravitational interaction at astrophysical scale, namely to appearance of
additional gravitational force, which can have an influence on movement of stars in galaxies
and galaxies in galactic clusters. The search of possible observational demonstrations of this
phenomenon is of direct physical interest.

It should be noted that discussed phenomena connected with the change of gravitational
interaction have essentially non-linear origin. Because of non-linear character of gravitating
vacuum approximative analysis of GTRC based on investigation of linearized theory and
perturbations of gravitational field above Minkowski space-time \cite{h} has to be
re-examined. In particular this concerns the analysis of particle content of GTRC, where it
would be taken into account not only deviation of space-time metric in the vacuum from that
of Minkowski space-time, but also presence of the vacuum torsion (compare with \cite{ru})
\footnote{The applying of \cite{h} to minimum GTRC shows that weak gravitational field
(by neglecting terms with $\omega$ in gravitational equations) in the frame of this theory
besides massless graviton includes massive particles with spin-parity $2^+$.}.

\section{Acknowledgements}

Author is thankful to the participants of the plenary meeting of International Conference on
Gravitation, Cosmology and Astrophysics (Kaliningrad, Russia, June 2017), and also to the
participants of Seminar of Department of Theoretical Physics and Astrophysics of Belarusian
State University (Minsk, Belarus) for fruitful discussions.

\end{document}